Instability of localized pulses in nonlinear electrodynamics

M.B. Belonenko, N.N. Konobeeva
Volgograd State University, 400062, Volgograd, Russia
e-mail: yana_nn@inbox.ru

We analyse the development of instability in the framework of nonlinear electrodynamics based on the Maxwell's equations without approach of slowly varying amplitudes and phases. The action is chosen from the Heisenberg-Euler Lagrangian, based on invariants of the electromagnetic field. The resulting scenario for the development of instability is consistent with the previously made conclusion in the framework of the approximation of slowly varying amplitudes and phases.

## 1. Introduction

It is well-known that Maxwell's equations in vacuum at high electric and magnetic field become non-linear due to the exchange of virtual electron-positron pairs [1,2]. In the general case, this effect is described by the Heisenberg-Euler Lagrangian [3, 4], which allows to obtain an expression for the effective Lagrangian of the electromagnetic field in powers of the electromagnetic field invariants. The well-known Lagrangian of the classical electromagnetic field is only a first-order term and contains only one invariant (scalar). In this case, we have the natural question about studying the effects of higher orders, i.e. about nonlinear vacuum electrodynamics (NED), which follows from the formalism of quantum electrodynamics. It should be noted that a number of studies on this topic have been provided to the noment [5–8]. Generally, these studies based on the slowly varying amplitude and phase approximation (SVAP). For example, in [5] the questions of three-wave mixing and mode instabilities in a resonator were investigated. In [6] based on SVAP it was concluded that there are instability of light bullets in the framework of NED, i.e. light bullets either experience dispersive spreading (with a maximum amplitude approaching zero), or they collapse in a finite time. Also in this paper, there are estimates for the limiting power that determines a particular mode. Note that the SVAP has a number of fundamental flaws despite the technical simplicity. First of all, it is a change in the dispersion law of linear oscillations in the system of Maxwell`s equations before and after the SVAP application. Besides this, after using the SVAP method, both spatial and temporal dispersions of nonlinear terms are not taken into account. Such effects are well-known in nonlinear optics [9], as well as the consequences to which they lead. Accordingly, it is possible to supplement the equations obtained by the SVAP with additional terms, and try to find solution of the system of Maxwell`s equations without applying the SVAP.

The second variant is also supported by the fact that Maxwell's equations can be reduced to a system of two wave equations. While the effective equation obtained by SVAP is a modification of the nonlinear Schrödinger equation for which explicit numerical schemes are not applicable. All these fact determined the subject of this work.

## 2. Basic equations

The system of Maxwell`s equations (with explicit vectors **D** and **H** through vectors **E** and **B**) has the following form:

$$\begin{aligned}
&\nabla\cdot\boldsymbol{E}=\left(\rho-\nabla\cdot\boldsymbol{P}\right)/\varepsilon_0,\\
&\nabla\cdot\boldsymbol{B}=0,\\
&\frac{\partial\boldsymbol{B}}{\partial t}+\nabla\times\boldsymbol{E}=0,\\
&\frac{1}{c^2}\frac{\partial\boldsymbol{E}}{\partial t}-\nabla\times\boldsymbol{E}=-\mu_0\left(\boldsymbol{j}+\frac{\partial\boldsymbol{P}}{\partial t}+\nabla\times\boldsymbol{M}\right)
\end{aligned}\tag{1}$$

This system, in the absence of free charges and currents, can be written as:

$$\begin{aligned}
&\frac{1}{c^2}\frac{\partial^2\boldsymbol{E}}{\partial t^2}-\nabla^2\boldsymbol{E}=-\mu_0\left(\frac{\partial^2\boldsymbol{P}}{\partial t^2}+c^2\nabla\left(\nabla\cdot\boldsymbol{P}\right)+\frac{\partial}{\partial t}\left(\nabla\times\boldsymbol{M}\right)\right),\\
&\frac{1}{c^2}\frac{\partial^2\boldsymbol{B}}{\partial t^2}-\nabla^2\boldsymbol{B}=\mu_0\left(\nabla\times\left(\nabla\times\boldsymbol{M}\right)+\frac{\partial}{\partial t}\left(\nabla\times\boldsymbol{P}\right)\right)
\end{aligned}\tag{2}$$

To establish the relationship of the vectors **M** and **P** with the vectors **E** and **B** we write the well-known Lagrangian Heisenberg-Euler [10-12]

$$L_c=-\frac{\alpha}{2\pi}\varepsilon_0E_{crit}^2\int_0^{i\infty}\frac{dz}{z^3}e^{-z}\left[z^2\frac{ab}{E_{crit}^2}coth\left(\frac{az}{E_{crit}}\right)cot\left(\frac{bz}{E_{crit}}\right)-\frac{z^2\left(a^2-b^2\right)}{3E_{crit}^2}-1\right]\tag{3}$$

where

$$a=\left[\left(F^2+G^2\right)^{0.5}+F\right]^{0.5},b=\left[\left(F^2+G^2\right)^{0.5}-F\right]^{0.5},\tag{4}$$

moreover, the values $a$ and $b$ are associated with scalar and pseudoscalar invariants of the electromagnetic field, which are given as:

$$\begin{aligned}
&F\equiv\frac{1}{4}F_{ab}F^{ab}=\frac{1}{2}\left(c^2\boldsymbol{B}^2-\boldsymbol{E}^2\right),G\equiv\frac{1}{4}F_{ab}\hat{F}^{ab}=-c\,\boldsymbol{E}\cdot\boldsymbol{B},\\
&\hat{F}^{ab}=\varepsilon^{abcd}F_{cd}/2
\end{aligned}\tag{5}$$

and $g_{ab}=diag(-1,1,1,1)$.

Decomposing (3) in a series and leaving the first two terms, we obtain for an effective Lagrangian the following expression:

$$L = L_0 + L_c = \varepsilon_0 F + \varepsilon_0^2 \kappa \left(4F^2 + 7G^2\right),$$
$$\kappa \equiv \frac{2\alpha^2 \hbar^3}{45 m_e^4 c^5} = \frac{\alpha}{90\pi}\frac{1}{\varepsilon_0 E_{crit}^2} \approx \frac{1}{3\times 10^{29}\,\mathrm{J/m^3}} \tag{6}$$

Taking into account that $\boldsymbol{P} = \delta L_c / \delta \boldsymbol{E}$, $\boldsymbol{M} = \delta L_c / \delta \boldsymbol{B}$ it is easy to obtain the expression for magnetization and polarization:

$$\boldsymbol{P} = 2\kappa\varepsilon_0^2 \left[ 2\left(E^2 - c^2 B^2\right)\boldsymbol{E} + 7c^2 \left(\boldsymbol{E}\cdot\boldsymbol{B}\right)\boldsymbol{B} \right],$$
$$\boldsymbol{M} = 2\kappa\varepsilon_0^2 c^2 \left[ -2\left(E^2 - c^2 B^2\right)\boldsymbol{B} + 7\left(\boldsymbol{E}\cdot\boldsymbol{B}\right)\boldsymbol{E} \right] \tag{7}$$

We assume, as usual, that the field amplitude is less than the critical one and the frequency is still quite small:

$$\omega \ll \omega_e \equiv m_e c^2 / \hbar, \left|\boldsymbol{E}\right| \ll E_{crit},$$
$$E_{crit} = \frac{m_e c^2}{e\lambda_e} \sim 10^{18}\,\mathrm{V/m} \tag{8}$$

The effects associated with the frequency increasing can be taken into account with the adding to the Lagrangian (6) terms in the following form [13]:

$$L_D = \sigma\varepsilon_0 \left[ \left(\partial_a F^{ab}\right)\left(\partial_c F_b^c\right) - F_{ab} \Box F^{ab} \right],$$
$$\Box = \partial_a \partial^a, \sigma = \left(2/15\right)\alpha c^2 / \omega_e^2 \approx 1.4\times 10^{-28}\,\mathrm{m}^2 \tag{9}$$

It is also possible to take into account the following terms in the expansion in invariants [14]:

$$L_3 \simeq \frac{2\alpha}{315\pi}\frac{e^4}{m_e^8} F\left(8F^2 + 13G^2\right) \tag{10}$$

A discussion of the possible consequences of such replacement is given below. The system of equations (2, 7) is the starting point for numerical simulation.

## 3. Main results of numerical simulation.

We choose the initial conditions for the system of equations (2, 7) in the form:

$$E_z = 0,$$
$$B_x = B_y = B_z = 0,$$
$$E_x = E_y = A\cdot exp\left(-\left(\frac{z}{\gamma}\right)^2\right) exp\left(-\frac{x^2+y^2}{\gamma_p^2}\right), \tag{11}$$
$$\frac{d}{dt}E_z = \frac{d}{dt}B_x = \frac{d}{dt}B_y = \frac{d}{dt}B_z = 0,$$
$$\frac{d}{dt}E_x = \frac{d}{dt}E_y = \frac{2vA}{\gamma^2}\cdot exp\left(-\left(\frac{z}{\gamma}\right)^2\right) exp\left(-\frac{x^2+y^2}{\gamma_p^2}\right)$$

This corresponds to the propagation at the initial moment of a cylindrically symmetric pulse of the electric field in a Gaussian form. In Eq. (11) $v$ is the initial velocity of the pulse, $\gamma$ is the width of the pulse along the propagation direction, $\gamma_p$ is the width of the pulse in the direction perpendicular to the propagation direction. Subsequently, we make a transition to the cylindrical coordinate system:

$$
\begin{aligned}
&\Delta\bar{B} = \left(\Delta B_\rho - \frac{B_\rho}{\rho^2}\right)\hat{\rho} + \left(\Delta B_\varphi - \frac{B_\varphi}{\rho^2}\right)\hat{\varphi} + \Delta B_z \cdot \hat{z}, \\
&\bar{\nabla}\times\bar{B} = -\frac{\partial B_\varphi}{\partial z}\hat{\rho} + \left(\frac{\partial B_\rho}{\partial z} - \frac{\partial B_z}{\partial \rho}\right)\hat{\varphi} + \frac{1}{\rho}\frac{\partial\left(\rho B_\varphi\right)}{\partial \rho}\hat{z}, \\
&\bar{\nabla}\cdot\bar{B} = \frac{1}{\rho}\frac{\partial\left(\rho B_\rho\right)}{\partial\rho} + \frac{\partial B_z}{\partial z}, \\
&\Delta f = \frac{1}{\rho}\left(\frac{\partial}{\partial\rho}\left(\rho\frac{\partial f}{\partial\rho}\right)\right) + \frac{\partial^2 f}{\partial z^2}, \\
&\bar{\nabla} f = \frac{\partial f}{\partial\rho}\cdot\hat{\rho} + \frac{\partial f}{\partial z}\cdot\hat{z}, \\
&\rho = \sqrt{x^2+y^2},\, tg\varphi = \frac{y}{x}
\end{aligned} \tag{12}
$$

and we apply the cross type numerical scheme [15].

The typical evolution of the pulse is shown in Figure 1.

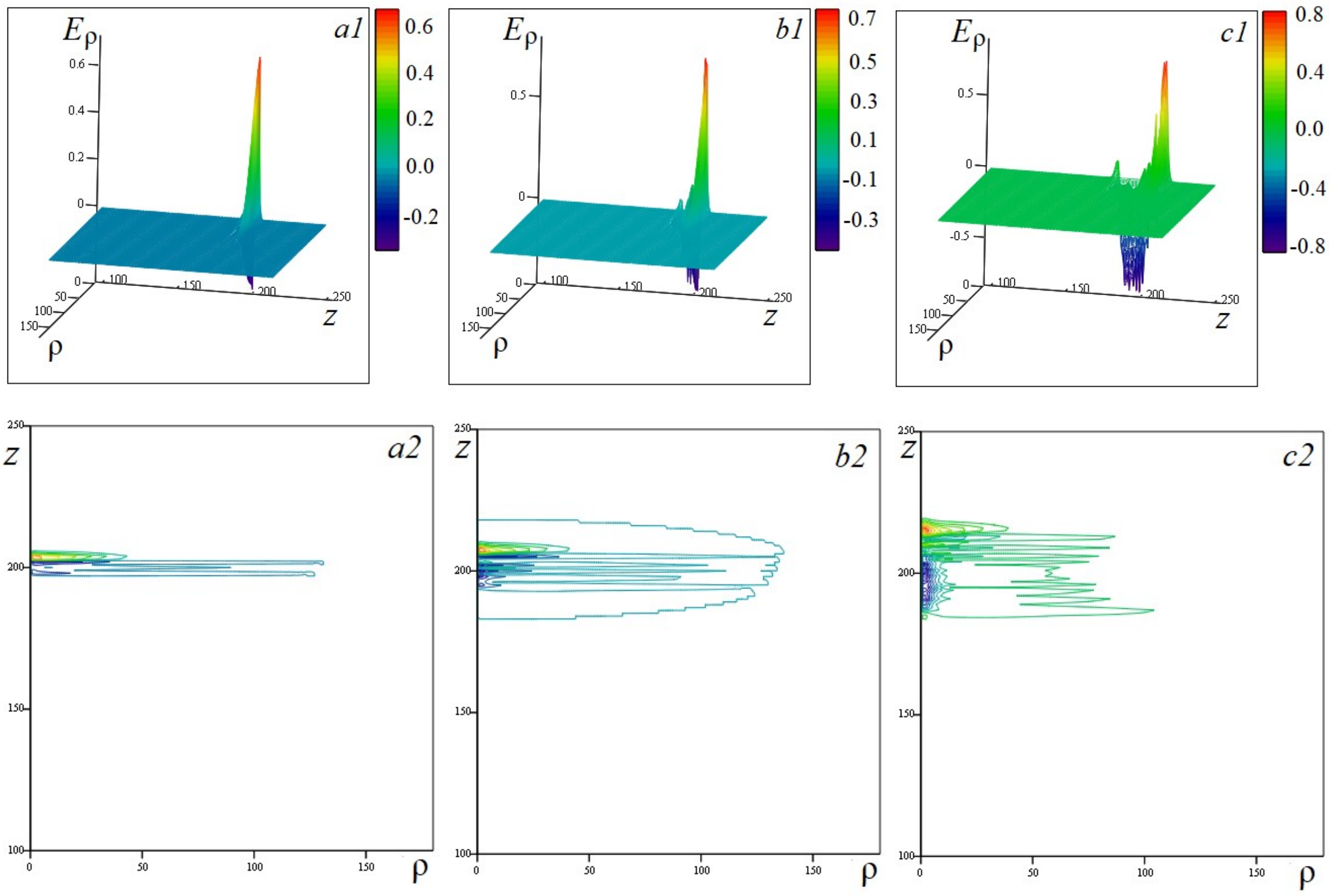

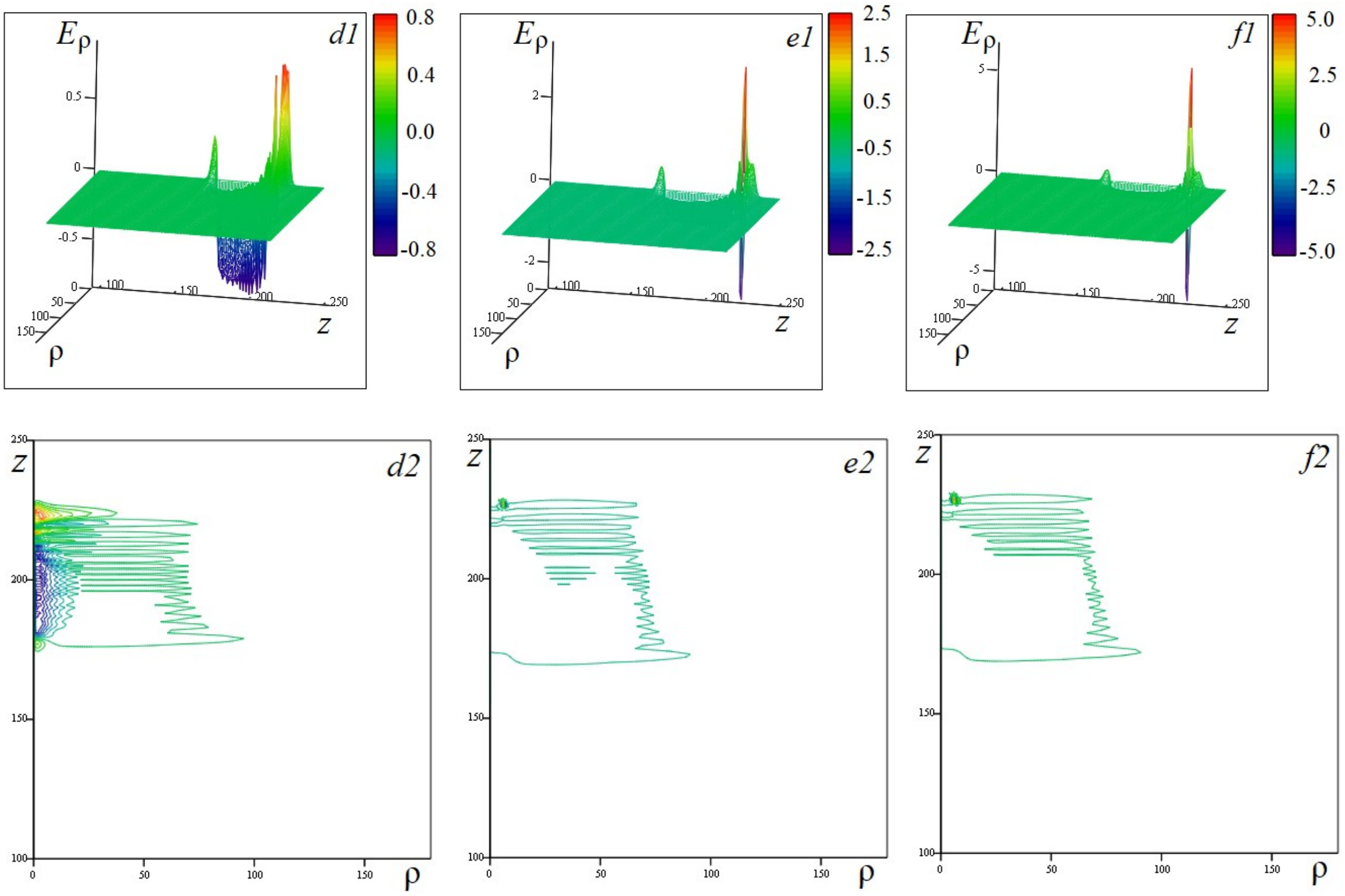

Fig. 1. The dependence of the electric field strength component $E_\rho$ on the coordinates at different time points: (*a*1, *a*2) *t*=50; (*b*1, *b*2) *t*=100; (*c*1, *c*2) *t*=200; (*d*1, *d*2) *t*=300; (*e*1, *e*2) *t*=380; (*f*1, *f*2) *t*=385. A three-dimensional view (numbers of figures with index 1) and lines under it for each time point (figures with index 2) are shown. All values are in relative units.

And, as usual, $E_\rho = E_x \cdot \cos\varphi + E_y \cdot \sin\varphi$. Amplitude slices $E_\rho$ on the z-axis are presented in Figure 2.

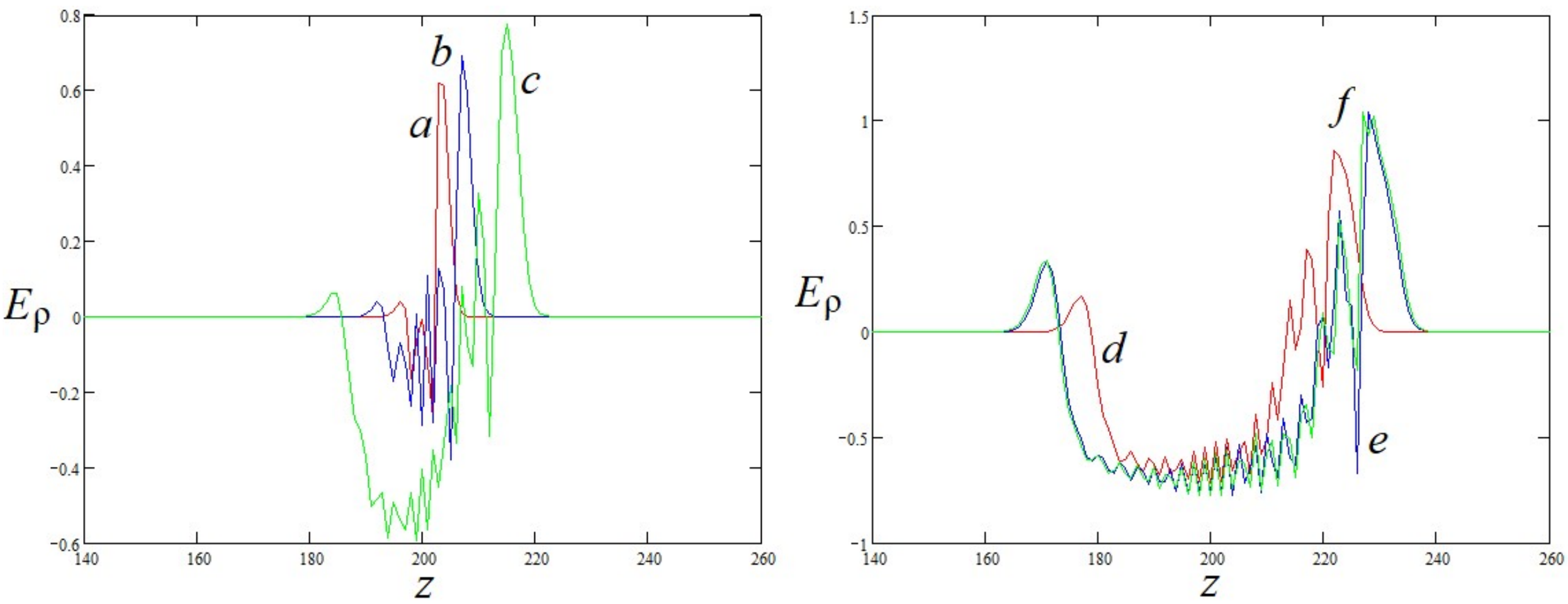

Fig. 2. The dependence of the electric field strength component $E_\rho$ for longitudinal section on the coordinates at different time points: a) *t*=50; b) *t*=100; c) *t*=200; d) *t*=300; e) *t*=380; f) *t*=385. All values are in relative units.

Thus, it is possible to analyze the mechanism of instability arising from a powerful Gaussian pulse propagation in a vacuum. Note that in Ref. [5] there are only the estimate obtained by the variational method and using the SVAP for the effective pulse width and the conclusion about the collapse.

At the first stage of the extremely short optical pulse evolution within the framework of nonlinear electrodynamics, it divides into two parts, which move in opposite directions. This effect is usually not monitored as with SVAP, and, especially, with the unidirectional propagation approximation [16]. Further, on the pulse front, it splits into a series of smaller pulses, which are accompanied by diffraction spreading in the perpendicular direction. Like the previous effect, this was not detected earlier, because the variational method most often used for analysis contains an assumption about the preservation of the pulse shape (that is, only the parameters describing this form change over time).

The third stage is characterized by an exponential growth of the pulse amplitude at the front, i.e. its collapse. It should be noted, that the time, at which the third stage occurs, depends most strongly on the pulse amplitude at the initial moment. What is important here is that in the study of such powerful extremely short pulses, the pulse will rather quickly cease to satisfy the condition $\left|\vec{E}\right| \ll E_{crit}$ (see (8)) and the production of electron-positron pairs will begin, which is not taken into account in our model. The calculations show that the scenario described above is carried out in a wide range of initial pulse amplitudes and widths. Also, note that a decrease in the initial amplitude significantly increases the time until the third stage of the pulse evolution.

## 4. Discussion of the results

The main result is the scenario for the onset of the collapse of the extremely short electromagnetic field pulse within nonlinear electrodynamics. This scenario includes both “splitting” into several pulses, and a sharp increase in the first pulse in the propagation direction in a finite period of time. Within the framework of the nonlinear electrodynamics, other corrections to the Lagrangian (6) are possible.

So, you can take into account the amendment of the next order in invariants (10) and/or the amendment associated with a decrease in the pulse duration (9). Primary, we discuss the following order correction for invariants (10). This amendment leads to nonlinearities of the 5th order in fields **E** and **B** in (7). The importance of taking into account these nonlinearities follows, for example, from Ref. [17], where it is shown, that the corresponding constants can be increased by taking into account the contribution of quantum chromodynamics. It is most important, these nonlinearities have the same sign as the nonlinearities in (7) [18], i.e. this amendment only enhances the effect of these nonlinearities and leads to time decrease until the pulse reaches the collapse stage.

Calculations with amendments, provisions (10) are presented. They show the same results as described above (which is why they are not given in the text). The only difference is a significant reduction in time (depending on the initial pulse amplitude) up to several times. A completely different character is of a corrective, associated with the pulse duration increasing (9).

This term changes the fundamental dispersion relation of the linear problem and, in some approximations [19], leads to an effective equation in the SVAP approximation, which has the form of a nonlinear Schrödinger equation. The question of the right introduction of such a correction is, according to Ref. [19], quite complicated, since its simple inclusion in the equations leads to the appearance of an additional nonphysical branch of the dispersion curve.

At the same time, the appearance of the effective nonlinear Schrödinger equation, and, accordingly, of solitons described by it in the framework of such model in the SVAP approximation, gives us a hope for the appearance of stable solutions based on the solution of the full system of the Maxwell equations. This question needs additional research and will be considered in the following papers.

**Acknowledgment**

This work is carried out within the framework of the state assignment of the Ministry of Education and Science of the Russian Federation (project no. 2.852.2017/4.6).